# SOBRE A VIAGEM DE ENRICO FERMI AO BRASIL EM 1934

## Francisco Caruso[1,2,3,4] & Adílio Jorge Marques[5,6]

[1] Doutor em Física, Pesquisador Titular do Centro Brasileiro de Pesquisas Físicas.

[2] Professor Associado no Instituto de Física da Universidade do Estado do Rio de Janeiro

[3] Pesquisador Colaborador do HCTE da Universidade Federal do Rio de Janeiro

[4] francisco.caruso@gmail.com

[5] Doutor em História da Ciência, Professor Adjunto do Instituto do Noroeste Fluminense de Educação Superior, Universidade Federal Fluminense.

[6] adiliojm@yahoo.com.br

**RESUMO**

Enrico Fermi foi um dos maiores físicos do século passado. Em 1934, fez uma série de palestras no Brasil. Convidado por Theodoro Ramos para trabalhar em São Paulo preferiu trabalhar nos EUA a partir de 1938. No entanto, indicou Gleb Wataghin – a quem Cesar Lattes se referia como "pai" da Física brasileira – para vir em seu lugar. Wataghin fez história no Brasil, tornando-se um dos primeiros catedráticos da futura Universidade de São Paulo. Assim, Fermi, além de ser um nome importante para a História das Ciências, ao indicar Wataghin, acabou deixando indelével marca na criação e institucionalização da pesquisa científica nacional. Apesar disso, sabe-se muito pouco sobre a viagem de Fermi ao Brasil. Esse trabalho reconstitui da forma mais ampla possível os passos do famoso físico italiano em nossas terras.

**PALAVRAS-CHAVE:** Enrico Fermi; Física; História da Ciência; Brasil.

**ABSTRACT**

Enrico Fermi was one of the greater physicists of the XX century. In 1934, he gave several lectures in Brazil. Invited by Theodoro Ramos to work in São Paulo, he preferred to stay in Rome and went to the USA in 1938. However, Fermi recommended Gleb Wataghin to come in his place. Wataghin made history in Brazil, becoming one of the first Professors of the future São Paulo University. Besides his relevance to the History of Science, Fermi eventually leaved an indelible mark on the creation and institutionalization of national scientific research due to the indication of Wataghin. Despite this fact, very little is known about Fermi's trip to Brazil. This work tries to reconstruct the fullest possible steps of the famous Italian physicist in our lands.

## 1. Introdução

Enrico Fermi (1901-1954), quando veio ao Brasil, esteve onde? Com quem se encontrou? Que conferências ele proferiu? Muito pouco se sabe sobre os detalhes dessa viagem, decorridos 80 anos. Nesse artigo, apresentamos o resultado parcial de uma pesquisa histórica que pretende lançar luz sobre essas perguntas.

Fermi foi um dos maiores físicos do século passado [FERMI, 1962; BERNARDINI & BONOLIS, 2004; BASSALO & CARUSO, 2013]. Recebeu o Prêmio Nobel de Física, em 1938, nominalmente pela identificação de novos elementos radioativos [FERMI, 1934] que se relaciona à descoberta das reações nucleares em nêutrons lentos, com a qual percebeu que prótons e nêutrons são estados quânticos de uma mesma partícula, como havia sugerido Werner Heisenberg (1901-1976). Na verdade, havia descoberto a fissão do núcleo.

Além dessa importante contribuição, havia descoberto, em 1926, uma nova estatística quântica [FERMI, 1926] para descrever um gás de partículas com *spin* semi-inteiro, sujeitas ao princípio de exclusão de Pauli, hoje conhecida como Estatística de Fermi-Dirac [POINTON, 1967]. Mais precisamente em 1934, ano em que fez uma série de palestras na Argentina, no Uruguai e no Brasil, Fermi desenvolveu sua famosa teoria do decaimento beta [FERMI, 1934a,b], uma das manifestações mais comuns da interação fraca. Na verdade, ele havia começado a esboçá-la no ano anterior [FERMI, 1933]. Seu artigo, recusado pela revista *Nature*, por ser considerado muito abstrato, foi publicado em uma revista italiana com o título "Tentativo di una nuova teoria dei raggi Beta", trabalho esse que levaria, quarenta anos depois, à Teoria Eletrofraca. Quanto ao título, cabe uma observação: ele deve ser entendido mais como uma figura de linguagem do que ao pé da letra, pois, como bem ressalva Nicola Cabibbo (1935-2010) [BERNARDINI & BONOLIS, 2004, p. 138],

> (...) a teoria apresentada naquele artigo é muito mais do que uma tentativa e com alguns aperfeiçoamentos (...) é perfeitamente válida hoje à distância de quase setenta anos.

Enrico Fermi, além de sua dedicação à pesquisa científica e de seu impressionante legado, também se preocupava em divulgar a ciência que fazia [FERMI, 2009] e teve um papel decisivo no desenvolvimento da Física no Brasil, embora indireto. De fato, esse papel poderia ter sido infinitamente maior caso ele tivesse aceitado o convite que lhe foi feito, em 1934, para vir trabalhar no Brasil, pelo matemático e político brasileiro Theodoro Augusto Ramos (1895-1937) que era professor na Escola Politécnica de São Paulo. Este havia sido comissionado pelo então governador de São Paulo, Armando de Salles Oliveira (1887-1945), para chefiar a comitiva acadêmica que foi à Europa, em abril de 1934, contratar pesquisadores[1] para a Faculdade de Filosofia, Ciências e Letras (FFCL), criada em 12 de janeiro deste ano, da qual Theodoro Ramos foi o primeiro Diretor [FERREIRA, 2012]. Importante ressaltar que a FFCL foi o núcleo da futura Universidade de São Paulo [CELESTE FILHO, 2009]. A ideia dessa comitiva nasceu da sugestão dada ao governador pelo matemático e acadêmico italiano Francesco Severi (1879-1961),[2] quando de sua estada em São Paulo, para que se criasse uma faculdade de Ciências [WATAGHIN, 1975] "que pudesse ser desenvolvida paralelamente às escolas profissionais nos moldes das universidades italianas." [SCHWARTZMAN, 1979].

Apesar do convite e insistência de Ramos, Fermi preferiu permanecer na capital italiana, onde havia sido nomeado, em 1927, Professor de Física Teórica da Universidade de Roma, cargo que ocupou até 1938, ano em que, logo depois de receber o Prêmio Nobel, imigrou para os EUA, onde inicialmente lecionou na Universidade de Columbia (BASSALO & CARUSO, 2013).

No entanto, para vir a São Paulo em seu lugar, Fermi indicou a Theodoro Ramos, o jovem físico ucraniano Gleb Vassielievich Wataghin (1899-1986) em quem "tinha confiança" e que na época estava na Universidade de Turim. Wataghin, que inicialmente recusou a indicação, acabou aceitando o convite [WATAGHIN, 1975] e vindo para o Brasil em abril de 1934,[3] onde fez história, tornando-se um dos primeiros catedráticos da futura Universidade de São Paulo, onde criou um importante grupo de pesquisas experimentais em raios cósmicos [FLEMING, 1996, p. 274] cujas pesquisas iniciais sobre a produção múltipla de mésons envolveu jovens cientistas como Marcelo Damy de Souza Santos (1914-2009) que foi aluno

da primeira turma para qual Wataghin deu aulas no Brasil. Depois veio Paulus Aulus Pompéia (1911-1993), que se tornou seu assistente, Oscar Sala (1922-2010), Yolande Monteux. Mais tarde, juntaram se ao grupo outros como, Giuseppe Occhialini (1907-1993), que veio da Itália à convite de Wataghin para trabalhar na USP de 1937 a 1944 e Cesar Lattes (1924-2005).

Wataghin também teve êxito em implantar a Física Teórica na USP, com a colaboração de Mario Schenberg (1914-1990), Abrahão de Moraes (1917-1970), Walter Schützer (?-1963), Paulo Saraiva de Toledo (1921-1999), e mais tarde com a vinda de Sonia Ashauer[4] (1923-1948) Jayme Tiomno (1920-2011), Paulo Leal Ferreira (1925-2005) e Roberto Salmeron (n. 1922). Wataghin providenciou a ida de alguns desses jovens para o exterior e esse foi um período muito importante para a consolidação futura da Física no Brasil. Mencionaremos aqui, brevemente, apenas um episódio que envolve Fermi diretamente. Wataghin havia enviado um trabalho teórico de Mario Schenberg a Dirac que decidiu convidá-lo para Cambridge. O jovem físico e seu mestre viajaram juntos e pararam em Roma, ocasião em que Schenberg foi apresentado a Fermi, que teria dito que não o deixaria sair de Roma e foi o que aconteceu. Schenberg ficou lá um ano e depois um ano com Pauli, voltando ao Brasil em 1938. Lattes gostava de se referir a Wataghin como "pai" da Física brasileira [LATTES, 2000, p. 35]. Salmeron refere-se a Wataghin com as seguintes palavras [SALMERON, 2002]:

> Não há muitos exemplos semelhantes de cientistas que, por suas ações pessoais, tiveram influência tão grande sobre tantas pessoas de gerações diferentes num país, que nem era o seu próprio país. Os alunos de Gleb Wataghin e os alunos de seus alunos espalharam-se por diferentes lugares, contribuindo para fazer da física brasileira o que ela é hoje.

Dessa forma, Enrico Fermi, além de ser um nome importante para a História das Ciências, ao indicar Wataghin, acabou deixando indelével marca na criação e institucionalização da pesquisa científica nacional. Apesar disso, sabe-se muito pouco sobre a viagem de Fermi ao Brasil. Nosso objetivo principal nesse trabalho é reconstruir, da forma mais ampla possível, os passos do famoso físico italiano em nossas terras e, para tal, em um primeiro, momento, utilizaremos as notícias publicadas em jornais da época.

**2. O trajeto de Fermi pelas notícias**

Laura Fermi, em seu livro de memórias [FERMI, 1954], dedica um capítulo de três páginas, intitulado *Interlúdio Sul-Americano*, à viagem que fez com seu marido no verão europeu de 1934 para cumprir uma agenda de palestras patrocinadas pelo governo italiano, vista como uma boa oportunidade para fugir do calor de Roma. Até onde sabemos, essa é a descrição mais completa dessa viagem. Ela nos dá o testemunho de que estiveram em Buenos Aires e Córdoba, na Argentina, em Montevidéu, no Uruguai, e em São Paulo e Rio de Janeiro, no Brasil. Todas as palestras foram dadas em italiano para plateias sempre muito cheias e demonstrando grande interesse pelo trabalho de Fermi [FERMI, 1954, pp. 94-95] e [SEGRE, 1970, p. 78].

As primeiras notícias que encontramos nos jornais da época sobre a vinda de Fermi ao Brasil datam de 10 de julho de 1934 e referem-se, principalmente, à criação do Instituto Ítalo-Brasileiro de Alta Cultura (IIBAC), instituição inaugurada por Massimo Bontempelli (1878-1960), acadêmico da Itália, em agosto de 1933. O fato se deu nas dependências da Academia Brasileira de Letras (ABL), contando com a presença de ilustres intelectuais brasileiros como Afrânio de Mello Franco (1870-1943), então Ministro das Relações Exteriores (*A Noite*, 1934, p. 2). Nota-se a preocupação do governo fascista de Mussolini de estabelecer laços acadêmicos e científicos com a América do Sul.

O mesmo Instituto, que seria depois instalado na *Casa d'Italia*, anuncia que virão ao Rio e a São Paulo, onde realizarão conferências, dois acadêmicos italianos: o eminente professor de fisiologia Fillippo Bottazzi (1867-1941), considerado o pai da Bioquímica na Itália, e o ilustre físico professor Enrico Fermi, cuja recente pesquisa o havia levado a prever a existência do "elemento 93" – isolado, em 1940, na Universidade da Califórnia, por Edwin Mattison McMillan (1907-1991) e Philip Hauge Abelson (1913-2004) –, despertou vivo interesse no mundo científico. O jornal *A Noite*, na mesma p. 2, noticia ainda que, provavelmente, virá também Giovanni Papini (1881-1956), e que o professor Carlos Chagas (1878-1934) será convidado a dar uma série de conferencias na Itália sobre doenças tropicais. Infelizmente, quatro meses depois, Chagas viria a falecer.

Em 24 de julho, reproduzindo uma notícia de Roma, da véspera, O *Jornal do Brasil*, p. 10, noticia que o acadêmico Enrico Fermi está a caminho de Buenos Aires onde realizará no Instituto Argentino de Cultura Italiana uma série de conferencias cientificas.

O Jornal *Correio da Manhã*, p. 7, de 26 de julho de 1934, traz a seguinte notícia sobre a descoberta de um novo elemento químico:

> Em uma das ultimas sessões da "Academia dei Lincei", em Roma, o jovem professor Enrico Fermi que conta apenas 32 anos de idade, annunciou haver descoberto um novo elemento chimico, cujas propriedades ainda está procurando pesquisar.

A notícia continua com uma descrição mais técnica da descoberta e termina dizendo que o nome do elemento será dado só após se conhecer suas principais propriedades.

Fermi e sua esposa, Laura Fermi (1907-1977), foram primeiro à Argentina, viajando a bordo do navio *Neptunia* (curiosamente o elemento sintético 93, inicialmente chamado de "ausônio",[5] por Fermi e seu Grupo de Roma [CORBINO, 1936], passaria a ser denominado "neptúnio"), proveniente de Trieste, que passou por Recife, Salvador e pela capital Rio de Janeiro em 26 de julho, onde deixou o tenor Aurelio Marcato que começava a despontar na Itália (*Jornal do Brasil*, p. 5, 27 de julho) e faria, em agosto, a "Maria Tudor" no Teatro Municipal do Rio de Janeiro (O Globo, p. 1, 26 de julho). O vapor zarpou, às 18 h, para seu destino final na capital do Rio da Prata, onde Fermi permaneceria 15 dias.

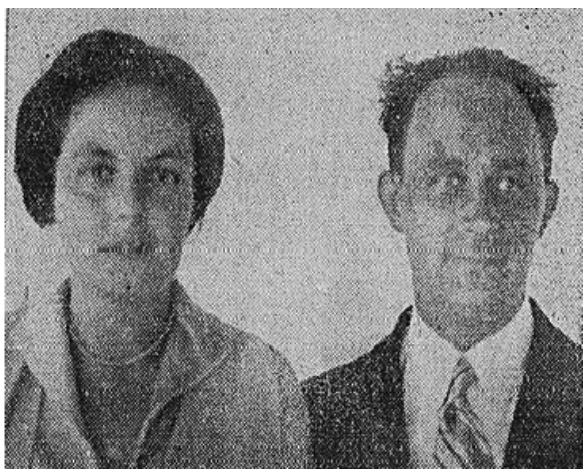

Figura 1. "O professor Enrico Fermi em companhia de sua senhora". *A Noite*, 3ª. Edição, quinta-feira, 26 de julho de 1934, primeira página.

A primeira página de *O Globo*, de 26 de julho, traz uma foto de Fermi e do tenor, assim como a lista dos passageiros que desembarcaram no Rio e menciona que Fermi deu uma palestra a bordo, uma espécie de síntese das conferências que faria em terra firme. De fato, em Buenos Aires, assim como faria depois no Brasil, daria uma série de conferências (*A Noite*, p. 1), num total de cinco (vermos, ao longo do texto, que foram quatro palestras no Brasil). Nesse mesmo ano, suas conferências na Argentina foram publicadas pela Faculdade de Ciências Exatas, Físicas e Naturais da Universidade de Buenos Aires [FERMI, 1934c], que havia sido fundada em 12 de agosto de 1821. Este fato, por si só, sugere uma grande diferença cultural e científica entre a Argentina e o Brasil da época, pois, até onde se sabe, nenhuma das palestras conferidas pelo físico italiano no Brasil fora transcrita. Nessa mesma manchete de *O Globo*, adianta-se a programação de palestras no Brasil:

> (...) a conferência em São Paulo, dia 22/08 e, no dia 25/08, às 17 h, na Academia Brasileira de Letras (ABL), onde falará sobre "A evolução da Physica no século XX". Será apresentado pelo acadêmico e professor Aloysio de Castro, Diretor brasileiro do IIBAC. No dia 28/08, às 17 horas, fará a segunda conferência no salão da ABL, "A produção artificial dos corpos radio-activos". No dia 1 de setembro volta para a Italia.

No dia 18 de agosto, o *Correio da Manhã*, p. 10, e o periódico *A Noite*, p. 9, noticiam a chegada de Fermi a São Paulo, prevista para o dia 22 deste mês. De fato, ele parte de Buenos Aires, a bordo do vapor *Northen Prince*, com destino ao Brasil (*Jornal do Brasil*, p. 10). Fermi desembarcou em Santos, dia 21 de agosto, e viajou de automóvel para a capital São Paulo, onde teve "honrosa recepção da parte das autoridades consulares, representantes do Governo e membros das sociedades científicas de S. Paulo" (*Jornal do Brasil*, p. 10, 22/08). Segundo a *Folha da Manhã*, de 22 de agosto, p. 10, Fermi hospedou-se no *Esplanada Hotel*. Este artigo traz uma pequena entrevista como físico italiano com suas impressões sobre as cidades de Santos e São Paulo. Essa palestra havia sido anunciada, em 20 de agosto, pela edição vespertina de *O Globo*, p. 2, e pelo *Correio de São Paulo*, com o título "A constituição da Matéria", na qual foi esperado que ele fizesse um contraponto entre as estatísticas de Einstein e da que ele próprio havia proposto recentemente. A *Folha da Manhã*, p. 6, e o *Correio Paulistano*, p. 4, confirmam, em 21 de agosto, que esta palestra ocorrerá dia 22 de agosto, às 21 h (veja também *Folha da Noite*, 22 de agosto, p. 4, 2ª. Edição), no Salão Nobre da sede do Instituto Histórico Geográfico em São Paulo, na Rua Benjamin Constant, 40, sob o patrocínio da USP, cuja fundação deu-se em janeiro de 1934. O clima em que essa palestra se deu é bem retratado na Folha da Manhã, de 23 de agosto, p. 4, que publica uma foto de Fermi cercado por várias pessoas com a legenda: "verdadeira multidão que encheu literalmente o amplo salão daquela entidade refluindo para seus corredores e salas". Ainda sobre o ambiente encontrado por Fermi, cabe o seguinte comentário. Embora sua visita tenha despertado um interesse geral não muito diferente da vinda de Einstein ao Brasil [TOLMASQUIM, 2003], em 1925, ela se dá em um momento histórico peculiar, no qual há uma série de iniciativas oficiais no sentido de criar um ambiente intelectual mais ativo e relevante no cenário nacional e, em particular, em São Paulo, com a criação da USP. Há, de fato, uma clara tentativa de dar os primeiros passos para a institucionalização da pesquisa científica no país, em diversas áreas do conhecimento. Uma medida da insipiência da produção científica no Brasil daquela época pode ser posta em evidência comparando-se dados disponibilizados por Joaquim Costa

Ribeiro, em "a Física no Brasil" [DE AZEVEDO, 1994, p. 207] com as obras completas de Fermi (1962). Conclui-se, assim, que o número total de trabalhos originais de Física publicados por brasileiros em periódicos científicos nacionais e estrangeiros, no biênio 1933-1934, é curiosamente idêntico ao número de trabalhos científicos publicados pelo próprio Fermi (12 no total), sendo que em 1933 o total de artigos brasileiros foi a metade do número de artigos de Fermi.

Em uma matéria do dia 23 de agosto, publicada no *Jornal do Brasil*, p. 13, baseada em informação de seu serviço especial em São Paulo, pelo telefone, encontramos a primeira entrevista (de um total de duas) dada por Fermi registrada nos jornais aos quais tivemos acesso, que esclarece o que será abordado em sua palestra. Transcrevemos a seguir suas palavras, pela primeira vez, fora da edição original:

> Limitar-me-ei a falar-vos em síntese. / Será tratado o problema das transformações dos elementos químicos um no outro, descrevendo as experiências feitas até hoje para a produção de novos corpos radiativos. Os métodos empregados – continuou o ilustre professor – para produzir essas transformações consistem em submeter a matéria a um bombardeio com partículas velocíssimas, que chocando os núcleos dos átomos, os modificam, transformando-os em núcleos de elementos diversos. A quantidade de matéria que se consegue transformar com esses métodos é pequeníssima e escaparia a qualquer método de análises químicas comuns, se não servisse para revelar o fato de que novos corpos formados são radioativos. Desintegrando-se, emitem elétrons velozes, dotados de energias de regular potência,[6] que permitem ser observados um por um. / Os novos corpos assim obtidos orçam em cerca de 40 e têm todos propriedades iguais ao rádio – concluiu o nosso entrevistado.

Essa interessante matéria traz, ainda, um pequeno depoimento do Físico e político italiano Orso Mario Corbino (1876-1937), grande amigo e incentivador de Fermi, apresentado pelo jornal como seu mestre, e que se refere ao elemento 93, descoberto por Fermi, com as seguintes palavras textuais:

> Trata-se do corpo simples mais duro que se conhece na atualidade. Esse elemento contém uma carga nuclear maior que qualquer outro e suas propriedades e radioatividades químicas são similares às do manganês. Com a aplicação do elemento 93, todos os materiais velhos poderão rejuvenescer.

Por fim, a reportagem divulga que, de 10 a 15 de setembro [1934], realizar-se-á em Veneza o Congresso Internacional de Rádio.

A primeira palestra de Fermi no Rio de Janeiro foi amplamente divulgada (*Correio da Manhã*, p. 7; *A Noite*, p. 2, ambos de 24/08, e o *Diário de Notícias*, p. 13, de 26/08; *O Globo*, edição vespertina, p. 2).

Tendo chegado ao Rio de Janeiro, capital, no dia 23 de agosto, o ilustre físico italiano foi recebido oficialmente na ABL, no Petit Trianon, na qualidade de "Acadêmico da Itália", no dia 25 do mesmo mês. Sua palestra, nessa ocasião, intitulada "A evolução da Física no século XX", teve repercussão em matéria publicada no *Jornal do Brasil*, p. 27, três dias depois, ou seja, ainda em 28 de agosto.

A sessão foi presidida por Cláudio [Justiniano] de Souza (1876-1954), que havia sido um dos fundadores da Academia Paulista de Letras, em 1909, e fundaria o Pen Clube do Brasil, em 1936. A presidência do evento designou dois membros-fundadores da ABL, Affonso Celso de Assis Figueiredo Júnior, o Conde de Affonso Celso (1860-1938) – filho do Visconde de Ouro Preto, último Presidente do Conselho de Ministros do Império – e Rodrigo Octavio de Langgaard Meneses (1866-1944), para introduzirem no recinto o professor Enrico Fermi, saudado por uma grande salva de palmas. O Médico e presidente da ABL, Aloysio de Castro (1881-1959), saudou-o em nome da Academia. Estiveram presentes personalidades como Celso Vieira de Matos Melo Pereira (1878-1954), Cláudio de Souza, Adelmar Tavares da Silva Cavalcanti (1888-1963), Affonso Celso, Antônio Joaquim Pereira da Silva (1876-1944), Rodrigo Octavio, Roberto Cantalupo (1891-1975), Embaixador da Itália, muitos Professores e numerosa assistência. Note-se que, excetuando-se o Embaixador, a imprensa só noticiou a presença de membros da ABL, não citando sequer um cientista.

Novamente na sede da ABL, Fermi realiza sua segunda conferência, no dia 28 de agosto, sobre a "A produção artificial dos corpos radio-activos" (texto de 26 de agosto no *Diário de Notícias*, p. 13). O Estado de São Paulo, de 29 de agosto, na matéria de capa, informa que, nessa ocasião, Enrico Fermi recebeu o diploma de membro da Academia Brasileira de Ciências,[7] sendo então saudado pelo acadêmico Professor Theodoro Ramos, já como membro honorário da Academia Brasileira de Letras. No dia seguinte, sai uma nota no *Jornal do Brasil*, p. 9, sobre a importância dessa série de palestras de Fermi sobre a "ciência nova" e o jornal reproduz um longo texto supostamente fiel ao que o jovem cientista italiano teria dito, como reproduzido no recorte a seguir.

A terceira e última palestra, também no Rio de Janeiro, será sobre o tema "Modelo Estatístico do Átomo e ocorrerá no Laboratório da Escola Politécnica (*O Globo,* p. 3, 29/08*; Jornal do Brasil*, p. 14, em 30 de agosto; *A Noite*, p. ilegível, de 31 do mesmo mês).

Em 14 de setembro, na edição matutina de *O Globo*, p. 2, lê-se a notícia, oriunda de Gênova, que, ao retornar de uma viagem ao Brasil e à Argentina, Fermi teria manifestado viva satisfação pela acolhida que teve na América Latina. Essa impressão foi compartilhada por Laura Fermi, que escreveu que a viagem foi um sucesso sob vários pontos de vista, principalmente pelo interesse manifestado pelo grande público nos trabalhos de seu marido [FERMI, 1954, p. 94].

Por fim, a Folha da Manhã, de 22 de outubro de 1934, na coluna "índices de alta cultura", trata do interesse e sucesso que as conferências de Fermi despertaram e conclui que a criação da USP correspondeu aos anseios de se buscar um novo patamar intelectual em São Paulo e ressalta a estabilidade da condição social e intelectual do povo, "apesar do agito político".

**3. Considerações finais**

Fruto de uma pesquisa ainda em andamento, esse é um breve registro da trajetória e das atividades de Enrico Fermi no Brasil, no curto período de tempo entre o dia 21 de agosto e primeiro de setembro de 1934, quando o cientista retorna à Itália. Essa história foi reconstituída quase que totalmente com base em textos de jornais da época o que, até onde sabemos, ainda não havia sido feito.

O material de imprensa ao qual tivemos acesso evidencia que, assim como ocorreu com Albert Einstein, em maio de 1925, a passagem e as atividades de Fermi pela América do Sul foram razoavelmente bem cobertas, embora há que se destacar que não no mesmo nível da visita de Einstein, muito bem contada em livro [TOLMASQUIM, 2003]. O que é curioso é que, ao contrário do caso do Einstein, essa visita de Fermi à América do Sul ficou praticamente ignorada por quase 80 anos, praticamente sem despertar a curiosidade das comunidades de físicos e de historiadores da ciência.

**Agradecimentos**

---

[1] O *Estado de São Paulo*, de 17 de maio de 1934, p. 2, noticia a chegada a Roma de Theodoro Ramos e informa que a USP contará com 10 catedráticos franceses, 4 italianos, 3 alemães, 1 inglês e 1 português. Em sua edição de 20 de maio do mesmo ano, o jornal divulga quem serão os cientistas alemães e que receberão um salário de 3,4 contos de réis. Em 2 de junho publica ainda o Estado de São Paulo, p. 4, uma entrevista detalhada de Theodoro Ramos quando de seu regresso ao Brasil.

[2] Há divergências na literatura acerca do nome do matemático procurado por Theodoro Ramos nessa ocasião. Segundo Simon Schwartzman, *op. cit.*, trata-se de Francesco Cerelli, enquanto para Patrick

Petitjean, em As Missões Universitárias Francesas na Criação da Universidade de São Paulo (1934–1940), in HAMBURGER, Amélia Império (Org.) *et. al. As Ciências nas Relações Brasil–França (1850–1950)*. São Paulo: USP/FAPESP, 1996, seria Francesco Severi. Esse nome é corroborado por Cesar Lattes, *in* Vera Maria de Carvalho e Vera Rita da Costa (Coords.), *Cientistas do Brasil: Depoimentos*, São Paulo: SBPC, 1998, p. 635. Provavelmente a origem dessa discordância seja o depoimento que Wataghin deu à Fundação Getúlio Vargas, *op. cit.*, pois, embora tenha participado da reunião, na página 11 da transcrição de sua entrevista refere-se ao matemático como Cerieli (*sic.*), enquanto se refere à mesma pessoa, na página 25, como Severi.

[3] Neste mesmo ano, outro núcleo de pesquisas físicas começou a aparecer no Rio de Janeiro, sob orientação de Bernhard Gross (1905-2002) [DE AZEVEDO, 1994, pp. 202-206] e Motoyama, "A Física no Brasil", in [FERRI & MOTOYAMA, 1979, pp. 71-73].

[4] Encaminhada por Wataghin, doutorou-se em 1948, na Universidade de Cambridge, sob a orientação de Paul Dirac (1902-1984), tornando-se a primeira mulher brasileira a concluir o doutorado em Física.

[5] Oscar D'Agostino (1901-1975) atribui a Franco Rasetti (1901-2001) a escolha do nome.

[6] A expressão "energia de regular potência" não faz sentido em Física.

[7] De fato, o tomo V, n. 3, dos *Anaes da Academia Brasileira de Sciencias*, de setembro de 1934, traz uma lista de 25 membros correspondentes na qual o nome Enrico Fermi aparece ao lado de outros como Albert Einstein (1879-1955), Jacques Hadamard (1865-1963), Irving Langmuir (1881-1957), Luiz Freire (1896-1963) e Bernhard Gross.

**Referências bibliográficas**

BASSALO, J.M.F. & CARUSO, F. *Fermi*. São Paulo: Livraria da Física, 2013.

BERNARDINI, C. & BONOLIS, L. (Eds.) *Enrico Fermi: His Work and Legacy*. Bologna: Società Italiana di Fisica; Berlin: Springer-Verlag, 2004.

CELESTE FILHO, M. Os primórdios da Universidade de São Paulo. *Revista Brasileira de História da Educação* v. 19, p. 187-204, 2009.

CORBINO, O.M. Il radio artificiale. L'ausonio e l'esperio. *Nuova Antologia* v. 71, p. 454-456, 1936.

D'AGOSTINO, O. *Il chimico dei fantasmi*. Atripalda: Mephite, 2002.

DE AZEVEDO, F. (Org.). *As Ciências no Brasil*, vol. 1. Rio de Janeiro: Editora da UFRJ, 1994.

DE MASI, D. (Org.) *A emoção e a Regra. Os Grupos Criativos na Europa de 1850 a 1950*. Rio de Janeiro: José Olympio Editora, 1999.

FERMI, E. Zur Quantelung des idealen einatomigen Gases. *Zeitschrift für Physik*, v. 36, p. 902-912, 1926.

FERMI, E. Tentativo di una teoria dell'emissione dei raggi «Beta». *Ricerca Scientifica*, v. 4, p. 491-495, 1933.

FERMI, E. Possible Production of Elements of Atomic Number High than 92. *Nature*, v. 133, p. 898-899, 1934.

FERMI, E. Tentativo di una nuova teoria dei raggi Beta. *Nuovo Cimento*, v. 11, p. 1-19, 1934(a).

FERMI, E. Versuch einer theorie der Beta-Stahlen. I. *Zeitschrift für Physik*, v. 88, p. 161-171, 1934(b).

FERMI, E. *Conferencias*. Faculdad de Ciencias Exactas, Fisicas y Naturales, Serie B, Publicacion 15, Buenos Aires, 1934(c).

FERMI, E. *The Collected Papers of Enrico Fermi*. 2 volumes. Chicago: University Press, 1962.

FERMI, E. *Atomi nuclei particelli. Scritti divulgativi ed espositivi 1923-1952*. A cura di Vincenzo Barone. Torino: Universale Bollati Boringhieri, 2009.

FERMI, L. *Atoms in the family. My life with Enrico Fermi*. Chicago: University Press, 1954.

FERREIRA, A. M. M. P. *A Criação da Faculdade de Filosofia, Ciências e Letras da Usp – Um Estudo Sobre o Início da Formação de Pesquisadores e Professores de Matemática e de Física em São Paulo*. Anais do 13º Seminário Nacional de História da Ciência e da Tecnologia, 2012, disponível em

http://www.sbhc.org.br/resources/anais/10/1344217546_ARQUIVO_TextoFinal-AlexandreM.M.P.Ferreira.pdf. Acesso em 4 de setembro de 2013.

FERRI, M.G. & MOTOYAMA, S. (Coord.). *História das Ciências no Brasil.* São Paulo: Editora Pedagógica e Universitária, 1979.

FLEMING, H. Enrico Fermi, gênio e simplicidade. *Revista Brasileira de Ensino de Física*, vol. 18, n. 4, p. 274-280, 1996.

*Hemeroteca Digital Brasileira*. http://hemerotecadigital.bn.br/. Acessos em 20 de julho de 2013.

LATTES, C. A Descoberta do Méson π, *in* G. Alves, F. Caruso, H. Motta & A. Santoro (Eds.), *O Mundo das Partículas de Ontem e de Hoje*. Rio de Janeiro: CBPF, 2000.

POINTON, A.J. *Introduction to Statistical Physics*. Bristol: Longmans, 1967.

ROCHA-FILHO, R.C. CHAGAS, A.P. Sobre os nomes dos elementos químicos, inclusive dos transférmios, *Química Nova*, v. 22, n. 5, p. 769-773, 1999.

SALMERON, R. Gleb Wataghin. *Estudos Avançados* v. 16, p. 310-315, 2002.

SCHWARTZMAN, S. *Formação da Comunidade Científica no Brasil*. São Paulo: Ed. Nacional, 1979.

SEGRÈ, E. *Enrico Fermi, physicist*. Chicago: University Press, 1970.

TOLMASQUIM, A.T. *Einstein – O Viajante da Relatividade na América do Sul*. Rio de Janeiro: Vieira & Lent, 2003.

WATAGHIN, G. *Gleb Wataghin (depoimento, 1975)*. Rio de Janeiro: CPDOC, 2010. Disponível em http://www.fgv.br/cpdoc/historal/arq/Entrevista477.pdf, acesso em 4 de setembro de 2013.